\documentclass[article,nojss]{jss}\usepackage[]{graphicx}\usepackage[]{xcolor}
\makeatletter
\def\maxwidth{ %
  \ifdim\Gin@nat@width>\linewidth
    \linewidth
  \else
    \Gin@nat@width
  \fi
}
\makeatother

\usepackage{alltt}

\usepackage{float}

\usepackage{subcaption}
\newcommand{\subfloat}[2][need a sub-caption]{\subcaptionbox{#1}{#2}}
\usepackage{amsmath}
\usepackage[ruled,vlined]{algorithm2e}
\usepackage{listings}
\usepackage[shortcuts]{extdash}
\usepackage{booktabs}
\usepackage{enumitem}

\def\cov{{\text{COV}}}

\def\var{{\text{var}}}

\setkeys{Gin}{width=0.45\textwidth}
\setcitestyle{square}

\newcommand{\fct}[1]{\code{#1()}}

\author{Ruoyong Xu\\University of Toronto
   \And Patrick Brown\\University of Toronto
   \And Pierre L’Ecuyer\\University of Montr\'eal}
\Plainauthor{Ruoyong Xu, Patrick Brown, Pierre L’Ecuyer}

\title{clrng: A tool set for parallel random number generation on GPUs in R}
\Plaintitle{}
\Shorttitle{}

\Abstract{
We introduce the \proglang{R} package \pkg{clrng}, which leverages the \pkg{gpuR} package and is able to generate random numbers in parallel on a Graphics Processing Unit (GPU) with the \pkg{clRNG} (OpenCL) library.  Parallel processing with GPU's can speed up computationally intensive tasks, which when combined with \proglang{R}, it can largely improve \proglang{R}’s downsides in terms of slow speed, memory usage and computation mode.  
Random number generation is essential for simulation-based statistical inference and modeling, yet there is currently no \proglang{R} package for GPU-based random number generation. \pkg{clrng} fills this gap. 
\pkg{clrng} enables reproducible research by setting random initial seeds for streams on both GPU and CPU, and can thus accelerate several types of statistical simulation and modelling. 
The random number generator in \pkg{clrng} guarantees independent parallel samples even when \proglang{R} is used interactively in an ad-hoc manner, such as in interrupted and restored sessions. This package is portable and flexible, developers can use its random number generation kernel for a wide range of applications and purposes.  
}

\Plainkeywords{GPU, clrng package, parallel computing, clRNG library}

\Address{
  Ruoyong Xu, Patrick Brown\\
  Department of Statistics\\
  University of Toronto\\
  700 University Ave., Toronto, ON M5G 1Z5, Canada\\
  E-mail: \email{ruoyong.xu@mail.utoronto.ca}, \email{patrick.brown@utoronto.ca}\\\\ 
  Pierre L’Ecuyer\\
  Department of Computer Science and Operations Research\\
  University of Montr\'eal\\
  Pavillon André-Aisenstadt, 
   2920 chemin de la Tour, 
   Montréal, QC,  Canada,  H3T 1J4\\
  E-mail: \email{lecuyer@iro.umontreal.ca}}
\IfFileExists{upquote.sty}{\usepackage{upquote}}{}
\begin{document}

\section[Introduction]{Introduction}

%

In recent years, parallel computing with \proglang{R} \citep{r2021} has become a very important topic and attracted lots of interest from researchers \citep[see][for a review]{eddelbuettel2021parallel}. Although \proglang{R} is one of the most popular statistical software with many advantages, it has drawbacks in memory usage and computation mode aspects \citep{zhao_2016}. To be more specific, (1) \proglang{R} requires all data to be loaded into the main memory (RAM) and thus can handle a very limited size of data; (2) \proglang{R} is a single-threaded program, it can not effectively use all the computing cores of multi-core processors. Parallel computing is the solution to these drawbacks, for an overview of current parallel computing approaches with \proglang{R}, see CRAN Task View by \citet{cran2021} at \url{https://cran.r-project.org/web/views/HighPerformanceComputing.html}. 

Graphics Processing Units (GPUs) have the potential to make an important contribution to parallel computing with \proglang{R}. GPUs can perform thousands of computations simultaneously, which makes them powerful for doing massively parallel computing, and they are relatively cheap compared to multicore CPU's. Although there have been a number of \proglang{R} packages developed which provide some GPU capability, they inevitably come with some limitations. Packages such as \pkg{gputools}, \pkg{gpumatrix}, \pkg{cudaBayesreg}, \pkg{rpud} (available on github), are no longer maintained, the popular \pkg{tensorflow} \citep{tensorflow1} package uses GPU via \proglang{Python}, which makes it difficult to include as a dependency for new \proglang{R} packages. All of these mentioned packages are restricted to \proglang{R} users with NVIDIA GPUs. \pkg{gpuR} \citep{gpur1} is the only \proglang{R} package with a convenient and flexible interface between \proglang{R} and GPUs, and it is compatible with many GPU devices. By utilizing the \pkg{ViennaCL} \citep*{rupp2016viennacl} library, it provides a bridge between \proglang{R} and non-proprietary GPUs through the \proglang{OpenCL} (Open Computing Language) backend, which when combined with \pkg{Rcpp} \citep{rcpp1} gives a building block for other \proglang{R} packages. 

Random number generation plays a crucial role in simulation-based statistical inference, machine learning, and various scientific fields. The quality of random numbers impacts the correctness of results of simulation studies.
While most random number generators are sequential, the \proglang{R} packages \pkg{parallel}, \pkg{future} \citep*{future1.19.1} and \pkg{rlecuyer} \citep{sevcikova2015package} are able to generate random numbers in parallel on multicore CPUs. More specifically, \pkg{parallel} writes an interface for the \pkg{RngStreams}, a \proglang{\texttt{C++}} library by \cite{l2002object} which is based on a combined multiple-recursive generator (MRG) MRG32k3a. \pkg{future} and \pkg{rlecuyer} also uses the combined MRG algorithm for generating random numbers. For up-to-date review papers on the generation of random numbers on parallel devices, and GPUs in particular, see: \cite{rLEC15a,rLEC17p,rLEC21a}.

The \pkg{clrng} package described here is currently the only \proglang{R} package that provides facilities for generating random numbers in parallel on GPU. Accomplishing this is complicated because each process must produce an independent (non-overlapping) sequence of random numbers, and in order to ensure reproducibility it should be possible to save and restore the current state of each stream of random numbers at any point.  By leveraging the \pkg{gpuR} package, \pkg{clrng} offers an extensible framework for \proglang{R} package developers to make use of these facilities in \proglang{C} or \proglang{OpenCL} code, and the random number generators guarantee independent parallel samples even when \proglang{R} is used interactively in an ad-hoc manner, allowing interrupted and restored sessions.

The remaining sections are organized as follows:
Section 2 introduces streams and the use of streams in work-items on a GPU device for generating uniform random numbers, and the usage of \fct{runifGpu}.
Section 3 presents two non-uniform RNGs in \pkg{clrng}, as examples for users to develop other RNGs of interest on GPUs in \proglang{R}.
Section 4 applies GPU-generated uniform random numbers in Monte Carlo simulation for Fisher’s exact test. This section explains how the random numbers are used, how the algorithm is parallelized and implemented on GPU, and provides two real data examples to demonstrate the function usage and its performance in \proglang{R}.
Section 5 showcases a useful application of Normal random numbers on GPU, it demonstrates the use of these numbers to simulate batches of Gaussian random surfaces with Mat\'ern covariance matrices simultaneously. This simulation also utilizes GPU-enabled functions from our other package \pkg{gpuBatchMatrix}.
Finally, the paper concludes with a short summary and a discussion in Section 6.

\section{Uniform random number generation} \label{}
Uniform random number generators (RNGs) are the foundation for simulating random numbers from all types of probability distributions. \cite{l2012random} summarized the usual two steps to generate a random variable in computational statistics: (1) generating independent and identically distributed (i.i.d.) uniform random variables on the interval $(0, 1)$, (2) applying transformations to these i.i.d. $U(0, 1)$ random variables to get samples from the desired distribution. \cite{l2012random} and \cite{robert2004random} present many general transformation methods for generating non-uniform random variables, for example, the most frequently used inverse transform method, the Box-Muller algorithm \citep{box1958note} for Gaussian random variable generation, and so on, which are all built on uniform random variables. 

\proglang{clRNG} \citep{l2015clrng} is an \proglang{OpenCL} library (available at \url{https://github.com/clMathLibraries/clRNG}) for uniform random number generation, it provides four different RNGs: the MRG31k3p, MRG32k3a, LFSR113, and Philox-4×32-10 generators. These four RNGs use different types of constructions. The \pkg{clrng} package uses the MRG31k3p RNG, making it able to generate random numbers on GPUs. We choose MRG31k3p as the base generator for the following reasons:  the original \pkg{RNGStreams} package \citep{l2002object} was built with MRG32k3a, which was designed to be implemented in double precision, and not with 32-bit integers. The MRG31k3p generator was designed later, specifically for 32-bit integer arithmetic, so it runs faster on the 32-bit GPUs. It is also faster than Philox-4×32-10. The MRG31k3p generator was introduced by \cite{rLEC00b}. MRG31k3p was also statistically tested extensively and successfully, \cite[see][]{rLEC07b}.

In what follows, we will illustrate how to create streams and how to use streams to generate uniform random numbers. 

\subsection{Creating streams}\label{createstreams}
When a RNG is called in parallel processes or successively called at several places in a program in \proglang{R}, random number generation would be more complicated because there is no guarantee that the streams (analogous to \code{.Random.seed} in \proglang{R}) do not overlap, and thus the generated sequences of random numbers may have dependence. \pkg{clrng} uses multiple distinct streams that are used in work-items that executes in parallel on a GPU device. A popular way of obtaining multiple streams is to take an RNG with a long period and cut the RNG's sequence into very long disjoint pieces of equal length $Z$, and use each piece as a separate stream. Creating a new stream amounts to computing its starting point. Each of these streams can also be partitioned into substreams with equally-spaced starting points \citep{l2002object, rLEC15a}, although this is not currently implemented in \pkg{clrng}. In general, a stream object contains three elements: the current state of the stream, the initial state of the stream (or seed). Streams are created sequentially in the way that whenever the user creates a new stream, the software automatically jumps ahead by $Z$ steps to find its initial state, and the two states in the stream object are set to it.

In the \proglang{clRNG} library, the MRG31k3p RNG's entire period of length approximately $2^{185}$ is divided into approximately $2^{51}$ non-overlapping streams of length $Z = 2^{134}$. 
The state (and seed) of each stream is a vector of six 31-bit integers. This size of state is appropriate for having streams running in work-items on GPU cards, while providing a sufficient period length for most applications. The initial state of the first stream (also called ``initial seed of the package'' or ``initial seed of the stream creator'' in \proglang{clRNG}) for the MRG31k3p RNG is by default $(12345, 12345, 12345, 12345, 12345, 12345)$.

\fct{setBaseCreator} sets the initial state of the first stream that is created, it plays a role in \pkg{clrng} like the function \fct{SetPackageSeed} in \pkg{RngStreams} and \fct{clrngSetBaseCreatorState} in \pkg{clRNG}. \pkg{clrng} is able to create streams both on the host and on the GPU device. The \fct{createStreamsCpu} function does the former. The \proglang{R} code below creates 4 streams on the host. 
\begin{CodeChunk}
\begin{CodeInput}
R> ## initializes the random number 
R> ## generator with the specified initial seed.
R> setBaseCreator(rep(12345,6))
R> 
R> ## creating 4 streams on CPU
R> myStreamsCpu <- createStreamsCpu(4)
R> t(myStreamsCpu)
\end{CodeInput}
\begin{CodeOutput}
##               [,1]       [,2]       [,3]       [,4]
## current.g1.1 12345  336690377  502033783  739421137
## current.g1.2 12345  597094797 1322587635 1475938232
## current.g1.3 12345 1245771585 1964121530  730262207
## current.g2.1 12345   85196284 1949818481 1630192198
## current.g2.2 12345  523477687 1607232546  324551134
## current.g2.3 12345 2094976052 1462898381  795289868
## initial.g1.1 12345  336690377  502033783  739421137
## initial.g1.2 12345  597094797 1322587635 1475938232
## initial.g1.3 12345 1245771585 1964121530  730262207
## initial.g2.1 12345   85196284 1949818481 1630192198
## initial.g2.2 12345  523477687 1607232546  324551134
## initial.g2.3 12345 2094976052 1462898381  795289868
\end{CodeOutput} 
\end{CodeChunk} 

\fct{setBaseCreator} takes a single argument, which is the initial state of the first stream, by default it is set to a vector of six 12345's.   
The argument of \fct{createStreamsCpu} is the number of streams to create. In this case, 4 streams are created, normally users would create many more than the 4 presented here, since most GPUs have thousands of work items. The default number of streams is set to the current number of total work items in use (i.e., \code{prod(getOption(`clrng.Nglobal')}).

We move streams to the GPU by converting them to a `\code{vclMatrix}'.
\begin{CodeChunk}
\begin{CodeInput}
R> myStreamsGpu = vclMatrix(myStreamsCpu)
\end{CodeInput} 
\end{CodeChunk} 

Equivalently, \fct{createStreamsGpu} creates streams directly on the GPU and returns a `\code{vclMatrix}', which makes it slightly more efficient when the number of streams is large.  

Here are a few important notes on the usage of these above functions.
\begin{itemize}
\item Calling \fct{setBaseCreator} establishes a hidden object \code{.Random.seed.clrng}, which (unlike the standard \code{.Random.seed}) controls the creation of new streams (not random numbers).   We allow users to call \fct{setBaseCreator}  at any time. \fct{setBaseCreator} should be called once before \fct{createStreamsCpu} and \fct{createStreamsGpu} if they would like to select their own initial seeds, or not at all, in which case streams take the package's default initial seed.  Otherwise, there is a slight probability of producing streams which overlap, particularly when an \proglang{R} session is restarted in the middle of the same program, and the user calls \fct{setBaseCreator} before creating new streams.

\item Users may switch between \fct{createStreamsCpu} and \fct{createStreamsGpu} in the same program, calling either will read and update \code{.Random.seed.clrng}. However, this is not recommended as it causes unnecessary data transfer between host and device.
\item If an \proglang{R} session is restarted during a task, as long as users saved the \proglang{R} environment (which Rstudio does automatically), the initial states (seeds) of the newly created streams will carry on from the last streams' seed from the previous \proglang{R} session.
\item Objects on the GPU are lost when an R session is restarted, even if the workspace is saved.  The object \code{myStreamsGpu} will remain in the workspace if the session is restarted, but the GPU memory containing \code{myStreamsGpu}'s data will no longer be accessible and an error will occur if the object is accessed.  To retain streams when a session is restarted the streams should be copied to the CPU using \code{as.matrix}.
\end{itemize}

Unlike \proglang{R}'s standard random number generators, the streams need to be explicitly specified when generating random numbers with \code{clrng}.  It would be possible to hide streams from the user by storing them as a hidden object, which the \pkg{rlecuyer} package does to some extent, we have chosen not to do so for two reasons.  First, the streams would need to be stored on the CPU and transferred to and from the GPU every time they were used.
Second, computations might be distributed across multiple GPU devices or multiple computers, and being explicit about which streams are used where ensures reproducibility and avoids the possibility that the same streams might be used on different devices without the user being aware.

\subsection{Generating uniform random numbers}

The streams created sequentially are used by work-items (the GPU analogy of CPU cores) on a GPU to generate random numbers. In \pkg{clrng} each work-item takes one distinct stream to ensure they don't interfere with each other's random number generation, so the number of streams created should always equal (or exceed) the maximum number of work items likely to be used. The main part of the kernel (functions for execution on the GPU) for generating uniform random numbers is shown in Listing \ref{lst:uniformkernel}. Kernels are written in the \proglang{C}-like \proglang{OpenCL} language, in which \code{__kernel} declares a function as a kernel, and the \code{__global} prefix to the pointer kernel arguments specifies that they point to global memory space accessible to all work items.  Users can set the argument \code{verbose=2} in the random number generator to print out the kernel, which is slightly more complex than the code in Listing \ref{lst:uniformkernel}. 
  
Here the pointers \code{streams} and \code{out} refer to the streams and the  output matrix respectively, which are stored in global memory. \code{Nrow} and \code{Ncol} represent row and column number of the matrix \code{out} respectively.   \code{Npad} is the ``internal'' number of columns, and \code{out} is an \code{Nrow} by \code{Ncol} submatrix of a larger \code{Nrow} by \code{Npad} matrix.

Each stream's current state is copied to the private memory of each work-item by the function \fct{streamsToPrivate}, in which \code{g1} and \code{g2} point to the first three and second three elements of stream states.
The object \code{startvalue} gives the position of the work item's stream in the \code{streams} object and is computed with the
\code{NpadStreams} object defined in a macro not shown here.   

The function \fct{clrngMrg31k3pNextState} generates an uniform random integer between 1 and 2147483647,  and then scaled to be in the interval $(0,1)$ by multiplying it by a constant \code{mrg31k3p\_NORM\_cl} (defined to be $1/2147483648$). If to generate random integers, \code{temp} is not scaled. At the end of generating random numbers, streams are transferred back to global memory through the function \fct{streamsFromPrivate}.

\begin{lstlisting}[language=C,basicstyle=\small,label={lst:uniformkernel},caption= Uniform random numbers generation kernel.]
__kernel void mrg31k3pMatrix(
  __global int* streams,
  __global float* out,
  int Nrow, int Ncol, int Npad){

int Drow, Dcol;
uint g1[3], g2[3];
double temp;
const int startvalue = (get_global_id(0) + 
get_global_size(0) * get_global_id(1)) * NpadStreams;

streamsToPrivate(streams,g1,g2,startvalue);

for(Drow = get_global_id(0); Drow < Nrow;
    Drow += get_global_size(0)){

    for(Dcol = get_global_id(1); Dcol < Ncol; 
        Dcol += get_global_size(1)){

        temp = fact * clrngMrg31k3pNextState(g1, g2);
        out[Drow * Npad + Dcol] = temp;

  }//Dcol
}//Drow

streamsFromPrivate(streams,g1,g2,startvalue);

}//kernel
\end{lstlisting}


Now we use the 4 streams created in section \ref{createstreams} to generate a vector of double-precision i.i.d. $U(0,1)$ random numbers with \fct{runifGpu}. To view the generated random numbers, we need to convert them to \proglang{R} vectors or matrices, by doing this, the random numbers are moved from the global memory of the GPU device to the host.
\begin{CodeChunk}
\begin{CodeInput}
R> getOption('clrng.type')
\end{CodeInput}
\begin{CodeOutput}
## [1] "double"
\end{CodeOutput}
\begin{CodeInput}
R> sim_1 =  runifGpu(n = 8, streams = myStreamsGpu, Nglobal = c(2,2))
R> as.vector(sim_1)
\end{CodeInput}
\begin{CodeOutput}
## [1] 0.735 0.842 0.614 0.216 0.110 0.870 0.649 0.170
\end{CodeOutput} 
\end{CodeChunk} 

The arguments of the \fct{runifGpu} are described as follows:
\begin{itemize}
\itemsep0em 
  \item \code{n}: either a scalar specifying the number of samples to generate or a vector of length 2 specifying the size of the output matrix.
  \item \code{streams}: streams for random number generation, which must be stored on the GPU as a \code{vclMatrix} object.
  \item \code{Nglobal}:  a (non-empty) vector specifying size of work items for use, with default value from the global option \code{clrng.Nglobal}.
  \item \code{type}: a character string specifying \code{double}, \code{float} or \code{integer} of random numbers, the default value is from the global option \code{clrng.type}. 
  \item \code{verbose}: a logical value, if \code{TRUE}, print extra information, default is \code{FALSE}.
\end{itemize}

The object returned is either a \code{vclMatrix} or \code{vclVector}, depending on whether \code{n} is two- or one-dimensional.  Reusing \code{myStreamsGpu} will produce a vector different from \code{sim_1}, as the current states of the streams in \code{myStreamsGpu} have advanced by two positions (each of the 4 work item generated 2 random numbers).  

As mentioned earlier, streams on the GPU do not remain in the memory when an \proglang{R} session is restarted. There are two ways to reproduce results, depending on how the program calls streams creator.
The simpler way to reproduce is by having a single program in an \proglang{R} script file with the initial seed for the creator set at the beginning.  But it is important that the program consistently creates the streams in exactly the same order. For example, like the toy \proglang{R} program provided below, we can reproduce the random matrix \code{sim_mat} just by keeping a record of the initial seed \code{c(11,22,33,44,55,66)}. Please note that this program serves as an illustration, as in practice, there is typically no need to constantly recreate new streams or replace existing ones every time we generate random numbers.

\begin{CodeChunk}
\begin{CodeInput}
R> setBaseCreator(c(11,22,33,44,55,66))
R> sim_mat <- matrix(0, nrow=10, ncol= 6)
R> for (i in 1:10){
R>   if( i %% 2 == 0){
R>     streams1 <- createStreamsGpu(4)
R>     sim_mat[i,] <- as.vector(clrng::rnormGpu(n=6, 
R>                              streams=streams1, Nglobal=c(2,2)))
R>   }else if(i==3){
R>     streams2 <- createStreamsGpu(4)
R>     sim_mat[i,] <- as.vector(runifGpu(n=6, 
R>                              streams=streams2, Nglobal=c(2,2)))
R>   }else{
R>     streams3 <- createStreamsGpu(8)
R>     sim_mat[i,] <- as.vector(rexpGpu(n=6, 
R>                              streams=streams3, Nglobal=c(4,2)))
R> }
R> }
R> sim_mat
\end{CodeInput} 
\end{CodeChunk}

Were we to replace the \code{for} loop above with a parallel equivalent (i.e.\ \code{mcmapply}), the order of stream creation could change with each program execution and the result would not be reproducible.  For more complicated applications, we recommend users to save the matrix of streams (current states and initial states) on the CPU in a file as a \code{.rds} object. And later recall the saved streams for regenerating results, the streams will start from their current states after they are loaded. This is a safer way than the previous one for reproducing results in simulations. Below we show the code that saves streams to a data file called \code{myStreams.rds} on CPU and then load it back and transfer it to a `vclMatrix' object \code{streams_saved} on GPU, and using it to (re)generate some Normal random numbers.
\begin{CodeChunk}
\begin{CodeInput}
R> saveRDS(as.matrix(myStreamsGpu), "myStreams.rds")
R> # Load the streams object as streams_saved
R> streams_saved <- vclMatrix(readRDS("myStreams.rds"))
R> clrng::runifGpu(n=6, streams=streams_saved, Nglobal=c(2,2))
\end{CodeInput} 
\end{CodeChunk}

\section{Some non-uniform random number generation}
\subsection{Normal random number generation}
We apply the Box-Muller transformation to $U(0,1)$ random numbers to generate standard normal random numbers. As shown in Algorithm \ref{algorithm1}, Box-Muller algorithm takes two independent, standard uniform random variables $U_1$ and $U_2$ and produces two independent, standard Gaussian random variables $X$ and $Y$, where $R$ and $\Theta$ are polar coordinate random variables. 
The Box-Muller algorithm is a very good choice for Gaussian transform on GPU's compared to other transform methods \citep{howes2007efficient}, because this algorithm has no branching or looping, which are the things GPU's are poor at. 

\begin{algorithm}[ht] 
\SetAlgoLined
 1, Generate $U_1$, $U_2$ i.i.d. from $U (0,1)$ \;
 2, Define \begin{align*} 
& R = \sqrt{-2*\log U_1},\\
&  \Theta = 2\pi*U_2,\\
&  X=R*\cos(\Theta),\\
& Y=R*\sin(\Theta);\;\end{align*}
 3, Take $X$ and $Y$ as two independent draws from $N(0,1)$\;
 \caption{Box-Muller algorithm.}
 \label{algorithm1}
\end{algorithm}

Listing \ref{lst:normal} shows a fragment of the kernel that generates standard Gaussian random numbers. The kernel has work items operating in pairs with shared local memory, the $U_1$ and $U_2$ are generated in parallel and stored in the two-dimenstional vector \code{part} below.  The \code{get_local_id(1)} command will return either zero or one, for the first and second item in the pair respectively.   As the work-items in a work-group proceed at different rates, \code{barrier(CLK_LOCAL_MEM_FENCE)} ensures correct ordering of memory operations to local memory, so that Gaussian random numbers $(X_1,Y_1), \dots, (X_n,Y_n)$ are generated in pairs correctly, errors such as $(X_n, Y_{(n-1)})$ or $(X_{(n-1)}, Y_{(n)})$ are avoided.

\begin{lstlisting}[language=C,basicstyle=\small,label={lst:normal},breaklines=true, escapeinside={(*@}{@*)}, caption = Normal random numbers generation kernel.]
__kernel void mrg31k3pMatrix(
  __global int* streams,
  __global double* out,
  int Nrow, int Ncol, int Npad, int NpadStreams){

int Drow, Dcol;
uint g1[3], g2[3];
int startvalue = (get_global_id(0) * get_global_size(1) + 
  get_global_id(1)) * NpadStreams;
const double fact[2] = { mrg31k3p_NORM_cl, TWOPI * mrg31k3p_NORM_cl };
const double addForSine[2] = { 0.0, - PI_2 };
local double part[2];

streamsToPrivate(streams,g1,g2, startvalue);

for(Drow=get_global_id(0); Drow < Nrow; Drow += get_global_size(0)){
  for(Dcol=get_global_id(1); Dcol < Ncol; Dcol += get_global_size(1)){
        
    part[get_local_id(1)] = fact[get_local_id(1)] * 
      clrngMrg31k3pNextState(g1, g2);
    barrier(CLK_LOCAL_MEM_FENCE);
    out[Drow * Npad + Dcol] = sqrt( -2.0*log(part[0]) ) * 
      cos(part[1] + addForSine[get_local_id(1)] );
    barrier(CLK_LOCAL_MEM_FENCE);
    
  }//Dcol
}//Drow
streamsFromPrivate(streams,g1,g2,startvalue);
}//kernel
\end{lstlisting}

Below we demonstrate the efficiency of \fct{rnormGpu} by generating a large-size matrix containing 100 million double-precision Gaussian random numbers, we compare the run-time of \fct{rnormGpu} against \fct{stats::rnorm}. Utilizing $512 \times 128$ work-items, \fct{rnormGpu} achieves a remarkable speed improvement, being more than 170 times faster than the standard (and single-threaded) \fct{stats::rnorm}. This performance gap widens as the matrix size increases.
\begin{CodeChunk}
\begin{CodeInput}
R> options('clrng.Nglobal' = c(512,128))
R> streams <- createStreamsGpu()
R> dim(streams)
\end{CodeInput}
\begin{CodeOutput}
## [1] 65536    12
\end{CodeOutput}
\begin{CodeInput}
R> system.time(rnormGpu(c(10000,10000), streams=streams))
\end{CodeInput}
\begin{CodeOutput}
##    user  system elapsed 
##   0.050   0.001   0.051
\end{CodeOutput} 
\end{CodeChunk} 

\begin{CodeChunk}
\begin{CodeInput}
R> system.time(matrix(rnorm(10000^2),10000,10000))
\end{CodeInput}
\begin{CodeOutput}
##    user  system elapsed 
##    6.70    1.11    7.80
\end{CodeOutput} 
\end{CodeChunk}

\subsection{Exponential random number generation}
The exponential random variates are produced by applying the inverse transform method on i.i.d. $U(0,1)$ random numbers. The random variable $X \sim \text{Exponential}(\lambda)$ has cumulative distribution function $F_{X}(x)=1-e^{-\lambda x}$ for $x \geq 0$ and $\lambda > 0$. The inverse of $F_{X}(\cdot)$ is $F^{-1}_{X}(y)= -(1/ \lambda) \log (1-y)$, for $ 0 \leq y <1$. The kernel for generating Exponential random numbers is similar to those for uniform and normal random numbers, except for the part that performs the inverse transform.

Below is an example that creates a $2 \times 4$ matrix of Exponential random numbers with expectation equal to 1.
\begin{CodeChunk}
\begin{CodeInput}
R> r_matrix <- rexpGpu(c(2,4), rate=1, myStreamsGpu, Nglobal=c(2,2))
R> as.matrix(r_matrix)
\end{CodeInput}
\begin{CodeOutput}
##      [,1]  [,2]  [,3]  [,4]
## [1,] 1.00 0.689 2.218 1.167
## [2,] 1.48 1.205 0.406 0.241
\end{CodeOutput} 
\end{CodeChunk} 

The arguments of the \fct{rexpGpu} are described as follows:
\begin{itemize}
\itemsep0em 
  \item \code{n}: a vector of length 2 specifying the row and column number if to create a matrix, or a number specifying the length if to create a vector.
  \item \code{streams}: streams for random number generation, streams cannot be missing.
  \item \code{Nglobal}: a (non-empty) integer vector specifying size of work items for use, with default value from the global option \code{clrng.Nglobal}.
  \item \code{type}: a character string specifying \code{double} or \code{float} of random numbers, the default value is from the global option \code{clrng.type}. 
  \item \code{verbose}: a logical value, if \code{TRUE}, print extra information, default is \code{FALSE}.
\end{itemize}

\section{An application of uniform RNG: Fisher's simulation}
The GPU-generated random numbers can be applied in suitable statistical simulations to accelerate computations.
One application of GPU-generated random numbers in \pkg{clrng} is Monte Carlo simulation for Fisher's exact test. Fisher’s exact test is applied for analyzing usually $2 \times 2$ contingency tables when one of the expected values in table is less than 5. Different from methods which rely on approximation, Fisher's exact test computes directly the probability of obtaining each possible combination of the data for the same row and column totals (marginal totals) as the observed table, and get the exact p-value by adding together all the probabilities of tables as extreme or more extreme than the one observed. However, when the observed table gets large in terms of sample size and table dimensions, the number of combinations of cell frequencies with the same marginal totals gets very large, \cite[][p. 23]{mehta2011ibm} shows a $5 \times 6$ observed table that has 1.6 billion possible tables. Calculating the exact P-values may lead to very long run-time and can sometimes exceed the memory limits of your computer. Hence, the option \code{simulate.p.value = TRUE} in \fct{stats::fisher.test} is provided, which enables computing p-values by Monte Carlo simulation for tables larger than 2 $\times$ 2. 

The test statistic calculated for each random table is $-\sum_{i,j}\log(n_{ij}!), i=(1,\dots, I), j=(1,\dots,J)$, (i.e., minus log-factorial of table), where $I$ and $J$ are the row and column number of the observed table. This test statistic can also be independently calculated for a table by \fct{clrng::logfactSum}. Given an observed table and a number of replicates $B$, the Monte Carlo simulation for Fisher's exact test does the following steps: 
\begin{enumerate}
\setlength\itemsep{0em}
  \item Calculate the test statistic for the observed table.
    \item In each iteration, simulate a random table with the same dimensions and marginal totals as the observed table using the \fct{rcont2} algorithm \citep[see][]{patefield1981efficient}, compute and optionally save the test statistic from the random table. 
    \item Count the number of iterations (\emph{Counts}) that have test statistics less or equal to the one from the observed table.
    \item Estimate p-value using $\frac{1 + \emph{Counts}}{B + 1}$.
\end{enumerate}

Step 1 and step 2-3 are done on a GPU with two kernels enqueued sequentially. For step 2, \fct{clrng::fisher.sim} adapted the function \fct{rcont2} used by \fct{stats::fisher.test} for constructing random two-way tables with given marginal totals. The algorithm samples the entries row by row, one at a time, conditional on the values of the entries already sampled. The conditional probabilities for the possible values of the next entry are updated dynamically each time a new entry is sampled. Then this entry is sampled by standard inversion of the cumulative distribution function, using one $U(0,1)$ random number.  For an $I \times J$ table, this requires $(I-1)(J-1)$ random numbers (the last column and last row do not need to be sampled). Finally, one can compute the test statistic for this newly sampled table. On a GPU, this step can be replicated say $n$ times in parallel by creating $n$ distinct random streams and launching $n$ separate work-items. Each work-item takes a random stream as input, performs all of Step 2 and 3, and returns the value of the test statistic on the GPU. Computing the p-value on the CPU (Step 4) is then a trivial operation.  Step 2 of saving test statistics from the random tables is made optional, which can reduce the run-time. By the way, there are a lot of other more recent methods for sampling (larger) contingency tables, many of them use Markov Chain Monte Carlo, the use of streams and GPU would be quite different in that case. See for examples \citep{10.1214/13-AOS1131, KAYIBI2018298, dyer1997sampling}. Doing the \fct{fisher.sim} function on GPU opens up many possibilities for future work, the specific implementation we've done for the tables isn't necessarily the optimal one. 

The arguments of the \fct{clrng::fisher.sim} are described as follows:
\begin{itemize}
\setlength{\parskip}{0pt}
  \item \code{x}: a contingency table, a vclMatrix of integers.
  \item \code{N}: an integer specifying number of replicates.
  \item \code{streams}: a vclMatrix of streams, streams cannot be missing.
  \item \code{type}: a character string specifying \code{double} or \code{float} of the returned test statistics, with default value from the global option \code{clrng.type}.
  \item \code{returnStatistics}: a logical value, if \code{TRUE}, it returns test statistics, default is \code{FALSE}.
  \item \code{Nglobal}: a (non-empty) integer vector specifying size of the index space on GPU for use, with default value from the global option \code{clrng.Nglobal}.
\end{itemize}
The test statistics (if returned) are of class `htest', similar to R's \fct{fisher.test} function.
Users request $N$ number of replicates, but the actual number of replicates executed on the GPU is larger, calculated as \code{ceiling(N/prod(Nglobal))*prod(Nglobal)}. 
To demonstrate the advantage of \fct{clrng::fisher.sim}, we compute the p-values for two real data examples: one with a relatively large p-value and another with a very small p-value. We then compare the run-time with using \fct{stats::fisher.test} for each of the data sets on two computers: one with a very good CPU and an ordinary GPU, and the other equipped with an excellent GPU and an ordinary CPU. Below, we show the \proglang{R} outputs for testing on computer 2.

\subsection{Comparing run-time: Monthly birth anomalies}\label{fisher_month}
The 2-way contingency table in Table \ref{tab:monthdata} comes from the 2018 Natality public use data provided by the Centers for Disease Control and Prevention’s National Center for Health Statistics \citep{National2018}.  The 2018 natality data file can be downloaded at \url{https://www.cdc.gov/nchs/data_access/VitalStatsOnline.htm}. 
Table \ref{tab:month} is a $12 \times 12$ table displaying frequencies for congenital anomalies of the newborn by birth month in 2018 within the United States. The column variables of these two tables represent the twelve categories of congenital anomalies of the newborn: 1) Anencephaly; 2) Meningomyelocele/Spina bifida; 3) Cyanotic congenital heart disease; 4) Congenital diaphragmatic hernia; 5) Omphalocele; 6) Gastrochisis; 7) Limb reduction defect; 8) Cleft lip with or without cleft palate; 9) Cleft palate alone; 10) Down syndrome; 11) Suspected chromosomal disorder; and 12) Hypospadias.

\begin{table}
\centering
\caption{\label{tab:monthdata}Monthly counts of birth anomalies.\label{tab:month}}
\centering
\begin{tabular}[t]{lrrrrrrrrrrrr}
\toprule
  & Ane & Men & Cya & Her & Omp & Gas & Lim & Cle & Pal & Dow & Chr & Hyp\\
\midrule
Jan & 29 & 55 & 172 & 46 & 39 & 73 & 48 & 183 & 77 & 103 & 102 & 174\\
Feb & 25 & 45 & 175 & 35 & 31 & 55 & 34 & 142 & 81 & 115 & 100 & 180\\
Mar & 31 & 48 & 182 & 41 & 47 & 72 & 40 & 200 & 86 & 90 & 96 & 180\\
Apr & 34 & 45 & 186 & 36 & 32 & 75 & 42 & 173 & 56 & 87 & 90 & 193\\
May & 33 & 40 & 187 & 46 & 24 & 80 & 35 & 180 & 75 & 91 & 100 & 197\\
Jun & 34 & 48 & 189 & 35 & 33 & 75 & 45 & 154 & 74 & 102 & 100 & 182\\
Jul & 26 & 43 & 198 & 34 & 21 & 74 & 36 & 179 & 79 & 86 & 92 & 193\\
Aug & 24 & 41 & 189 & 44 & 43 & 62 & 48 & 183 & 88 & 109 & 94 & 194\\
Sept & 34 & 44 & 147 & 40 & 37 & 66 & 36 & 158 & 73 & 112 & 103 & 196\\
Oct & 25 & 43 & 207 & 45 & 31 & 65 & 49 & 181 & 77 & 108 & 115 & 220\\
Nov & 36 & 55 & 188 & 39 & 39 & 62 & 43 & 144 & 68 & 98 & 79 & 173\\
Dec & 23 & 48 & 196 & 31 & 31 & 71 & 31 & 177 & 86 & 86 & 73 & 156\\
\bottomrule
\end{tabular}
\end{table}

\begin{CodeChunk}
\begin{CodeInput}
R> # using GPU
R> month_gpu <- vclMatrix(month, type = "integer")
R> system.time(result_month <- 
+  clrng::fisher.sim(month_gpu, 1e6, 
+                    streams=streams,
+                    returnStatistics=TRUE))
\end{CodeInput}
\begin{CodeOutput}
##    user  system elapsed 
##   0.348   0.001   0.348
\end{CodeOutput} 
\end{CodeChunk} 

\begin{CodeChunk}
\begin{CodeInput}
R> unlist(result_month[c('p.value','threshold','simNum','counts')])
\end{CodeInput}
\begin{CodeOutput}
##     p.value   threshold      simNum      counts 
##       0.403  -47954.798 1048576.000  422618.000
\end{CodeOutput} 
\end{CodeChunk} 

We obtained $422618$ cases where the test statistics are below the observed threshold, based on an actual number of 1048576 simulations.
The p-value from \fct{clrng::fisher.sim} for this table is approximately $4.03\times 10^{-1}$, which closely aligns with the p-value from \fct{stats::fisher.test}. \fct{clrng::fisher.sim} completes in approximately 0.3 seconds, accounting for only 2.2\% of the total elapsed time taken by \fct{stats::fisher.test}.

\begin{CodeChunk}
\begin{CodeInput}
R> ## using CPU
R> system.time(result_monthcpu<-stats::fisher.test(month,
+            simulate.p.value = TRUE, B=result_month$simNum))
\end{CodeInput}
\begin{CodeOutput}
##    user  system elapsed 
##  14.603   0.037  14.633
\end{CodeOutput}
\begin{CodeInput}
R> result_monthcpu$p.value
\end{CodeInput}
\begin{CodeOutput}
## [1] 0.404
\end{CodeOutput} 
\end{CodeChunk}

\subsection{Comparing run-time: Day-of-week birth anomalies}\label{fisher_week}
Table \ref{tab:week} displays a $7 \times 12$ table illustrating the frequencies of congenital anomalies of newborns categorized by the day of the week of their birth in the United States during 2018.

\begin{table}
\centering
\caption{\label{tab:weekdata}Day-of-week birth anomaly data\label{tab:week}}
\centering
\begin{tabular}[t]{lrrrrrrrrrrrr}
\toprule
  & Ane & Men & Cya & Her & Omp & Gas & Lim & Cle & Pal & Dow & Chr & Hyp\\
\midrule
Mon & 30 & 34 & 173 & 37 & 23 & 80 & 49 & 191 & 83 & 122 & 109 & 216\\
Tue & 60 & 121 & 383 & 80 & 83 & 131 & 71 & 349 & 146 & 164 & 168 & 352\\
Wed & 51 & 106 & 417 & 92 & 73 & 145 & 72 & 333 & 136 & 179 & 196 & 351\\
Thu & 60 & 86 & 362 & 69 & 74 & 120 & 85 & 326 & 132 & 220 & 187 & 359\\
Fri & 52 & 94 & 347 & 87 & 59 & 123 & 68 & 323 & 145 & 170 & 166 & 345\\
Sat & 52 & 63 & 323 & 67 & 64 & 135 & 73 & 316 & 170 & 189 & 188 & 357\\
Sun & 49 & 51 & 211 & 40 & 32 & 96 & 69 & 216 & 108 & 143 & 130 & 258\\
\bottomrule
\end{tabular}
\end{table}

\begin{CodeChunk}
\begin{CodeInput}
R> # using GPU
R> week_GPU<-gpuR::vclMatrix(week,type="integer")
R> system.time(resultWeek<-clrng::fisher.sim(week_GPU, 1e7, 
+                                streams=streams,
+                                returnStatistics=TRUE))
\end{CodeInput}
\begin{CodeOutput}
##    user  system elapsed 
##   2.024   0.007   2.030
\end{CodeOutput}
\begin{CodeInput}
R> unlist(resultWeek[c('p.value','threshold','simNum','counts')])
\end{CodeInput}
\begin{CodeOutput}
##         p.value       threshold          simNum 
##        0.000125   -54989.556980 10027008.000000 
##          counts 
##     1255.000000
\end{CodeOutput} 
\end{CodeChunk}

\begin{CodeChunk}
\begin{CodeInput}
R> # using CPU
R> system.time(result_weekcpu<-fisher.test(week,
+            simulate.p.value = TRUE, B=resultWeek$simNum))
\end{CodeInput}
\begin{CodeOutput}
##    user  system elapsed 
##  87.447   0.412  94.577
\end{CodeOutput}
\begin{CodeInput}
R> result_weekcpu$p.value
\end{CodeInput}
\begin{CodeOutput}
## [1] 0.000127
\end{CodeOutput} 
\end{CodeChunk} 

The ``week'' table yields a significantly smaller p-value, approximately $1.27\times 10^{-4}$. Achieving a more precise p-value requires a larger number of simulations. With over ten million simulations, we obtain $1255$ cases and a p-value around $1.25\times 10^{-4}$ using \fct{clrng::fisher.sim}. In comparison, \fct{stats::fisher.test} takes about 88 to 94 seconds, while \fct{clrng::fisher.sim} completes in about 2 seconds. The time taken by \fct{clrng::fisher.sim} is reduced to approximately 2.2\% of \fct{stats::fisher.test}.

\subsection{A summary of the results}
We summarized the comparison results in Table \ref{tab:summary} and plotted the test statistics in Figure \ref{fig4}.
\begin{table}[H]
\centering
\caption{\label{tab:summarycompare}Summary of comparions of Fisher's test simulation on 2 devices. Computer 1 is equipped with CPU Intel Xenon W-2145 3.7Ghz and AMD Radeon VII. Computer 2 is equipped with VCPU Intel Xenon Skylake 2.5Ghz and VGPU Nvidia Tesla V100 \label{tab:summary}. Run-time may vary slightly on each execution.}
\centering
\begin{tabular}[t]{l|l|l|l|l|l}
\hline
\multicolumn{1}{c|}{ } & \multicolumn{2}{c|}{Computer 1} & \multicolumn{2}{c|}{Computer 2} & \multicolumn{1}{c}{ } \\
\cline{2-3} \cline{4-5}
B & Intel 2.5ghz & AMD Radeon & Intel 3.7ghz & NVIDIA V100 & Data\\
\hline
\multicolumn{6}{l}{\textbf{P-value}}\\
\hline
\hspace{1em}1M & 0.403804 & 0.403507 & 0.4035606 & 0.403507 & month\\
\hline
\hspace{1em}10M & 0.0001251 & 0.0001274 & 0.0001202 & 0.0001274 & week\\
\hline
\multicolumn{6}{l}{\textbf{Run-time}}\\
\hline
\hspace{1em}1M & 49.3 & 2.2 & 15 & 0.3 & month\\
\hline
\hspace{1em}10M & 327.3 & 10.2 & 90 & 2 & week\\
\hline
\end{tabular}
\end{table}

\begin{figure}[h]

{\centering \subfloat[Month data\label{fig:fighistMonth-1}]{\includegraphics[width=0.47\textwidth]{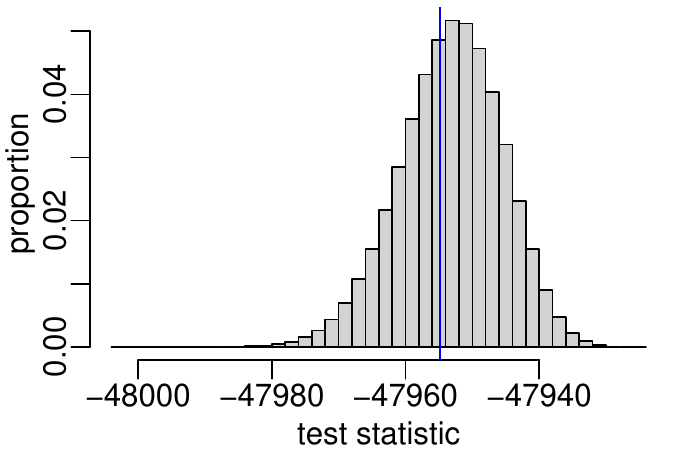} }
\subfloat[week data\label{fig:fighistMonth-2}]{\includegraphics[width=0.47\textwidth]{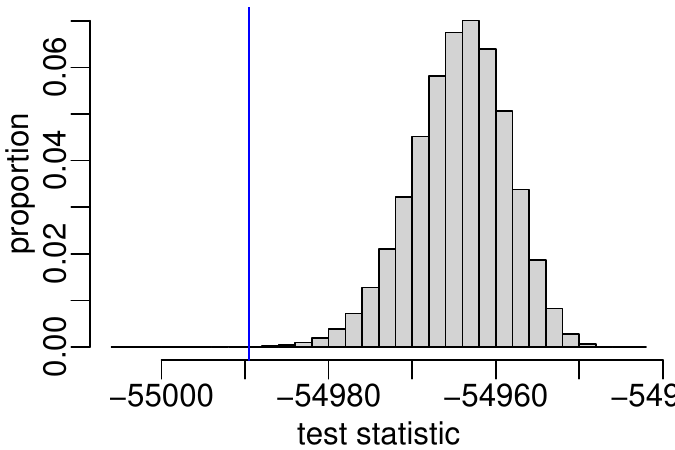} }

}

\caption[\label{fig4}Approximate sampling distributions of the test statistics from the two examples, Month and Week]{\label{fig4}Approximate sampling distributions of the test statistics from the two examples, Month and Week. The test statistic values of the observed tables are indicated by a blue vertical line on each plot.}\label{fig:fighistMonth}
\end{figure}

\section{Simulating Gaussian Random Fields} 
Simulating Gaussian random fields (GRFs) poses computational challenges due to the high dimensionality of the problem. For instance, a 100 by 100 grid results in 10,000 cells and a 10,000 by 10,000 covariance matrix. Writing $x_1, x_2, \dots, x_n$ as the point locations (e.g., centers of grid cells), the joint distribution of $U=(U_{x_1}, U_{x_2}, \dots, U_{x_n})^\top$ is a multivariate Normal (MVN) distribution.
An isotropic random field features a variance matrix with entries that depend solely on the distances between point locations. A GRF with geometric anisotoropy includes entries that depend on both the distances and directions between the points.
A geometrically anisotropic GRF with a Mat\'ern correlation function, as described by Mat\'ern (1960) \citep{matern1960spatial}, is defined as:
\begin{align*}
U &= [U(x_1), \ldots, U(x_n)]^\top \sim \text{MVN}(0, \Sigma), \\
\Sigma_{ij} &= \cov[ U(x_i),U(x_j) ]= \sigma^2 \frac{2^{\kappa-1}}{\Gamma(\kappa)} \left(\sqrt{8\kappa} ||d_{ij}||/\phi \right)^\kappa  K_\kappa\left(\sqrt{8\kappa} ||d_{ij}||/\phi \right),\\
d_{ij} &= 
        \left[
          \begin{array}{rr}
            1 & 0 \\
            0 & \omega
          \end{array}
        \right]
        \left[
          \begin{array}{rr}
            \cos(\theta) & -\sin(\theta) \\
            \sin(\theta) & \cos(\theta)
          \end{array}
        \right]
(x_i - x_j)
\end{align*}
where $K_\kappa(\cdot)$  is a modified Bessel function of the second kind with order $\kappa$ and $\Gamma(\cdot)$ is the standard gamma function.  The model parameters are:
\begin{itemize}
\item $\sigma$, the (marginal) variance  $\var(U_i) = \sigma$;
\item $\kappa$, the shape parameter controlling the differentiability of $U(\cdot)$;
\item $\phi$, the range parameter controlling how quickly correlation decays with distance in the direction where correlation is strongest;
\item $\omega \geq 1$, a parameter controlling how much faster correlation decays in the direction where correlation is weakest; and 
\item $\theta$, the angle of rotation  required to put the direction of maximum correlation on the x-axis.
\end{itemize}
The standard isotropic Mat\'ern model is obtained by setting $\omega = 1$.
There are several alternative parameterisations of Mat\'ern covariance functions in literature  \citep[see][]{haskard2007anisotropic}. In \pkg{gpuBatchMatrix}, we use the above form with the $\sqrt{8\kappa}$ term as it allows $\phi$ to be interpretable as the distance beyond which correlation is less than 0.14.
The conceptually simplest way of simulating a GRF is to multiply a vector of independent standard Normals by the Cholesky decomposition of $\Sigma$.  This method is known as the direct matrix decomposition method.  

\cite{LiuandLi2019} gives a comprehensive review on seven popular methods for GRF generation, 
all of these methods, except for the matrix decomposition method, involve approximations to the random field and have specific requirements on the type of grid or covariance functions. The direct matrix decomposition method is exact, it works for all covariance functions, can generate random fields on arbitrary point locations, and is straightforward to implement.

\begin{algorithm}[H]
\SetAlgoLined
 1, Calculate the covariance matrix $\Sigma$ between locations\;
 2, Compute the Cholesky decomposition of $\Sigma = L  D  L^\top$, where $L$ is a lower unit triangular matrix, and $D$ is a diagonal matrix\;
 3, Generate on GPU a random matrix $Z=(Z_1, Z_2, \dots, Z_n) \sim \text{MVN}(0,I_n)$, where $I_n$ is a $n \times n$ identity matrix\;
 4, Compute the random samples from $U$ in batches using $U = L  D^{\frac{1}{2}}  Z$ \;
 \caption{Gaussian random fields simulation using the direct matrix decomposition method.}
 \label{algorithm2}
\end{algorithm}

The direct method is detailed in Algorithm \ref{algorithm2}.  It is computationally demanding for two reasons.  First, evaluating the Bessel function is an iterative procedure that must be done for each of the $n^2/2$ entries of $\Sigma$.  Second, the time cost of Cholesky decomposition of the covariance matrix is $\mathcal{O}(n^3)$,  and is $\mathcal{O}(n^2)$ for the matrix-vector multiplication $L*Z$ \citep{LiuandLi2019}. Methods employing approximations gain benefits in computations while sacrificing exactness, we preserve the exactness and also largely reduce the computation burden by utilizing our other \proglang{R} package \pkg{gpuBatchMatrix}. \pkg{gpuBatchMatrix} address this issue since it computes batches of Mat\'ern covariance matrices in parallel, and does Cholesky decomposition and matrix-matrix multiplication in batches in parallel on GPU.

Other \proglang{R} packages that offer simulation of GRF's such as \pkg{geoR} \citep{geoR2001}, does not work for large number of locations. The \pkg{RandomFields} \citep{RandomFields2015} package has several different methods for simulation of Gaussian fields, among which the circulant embedding, which is an improved method on direct matrix decomposition, and some variants of the method like the cut-off embedding \citep{gneiting2006fast} are also exact and fast for isotropic GRF's, however, they work only on rectangular grids. 



\subsection{Simulating Gaussian random fields with Mat\'ern covariances}

Here is an example in which we simulate eight GRFs with Mat\'ern covariance on the GPU, using four sets of parameters simultaneously. This is achieved by leveraging the GPU capabilities provided by \pkg{clrng} and \pkg{gpuBatchMatrix} together. Motivated by the classic Swiss rain dataset, we simulate random fields on a grid of points covering Switzerland.

Step 1, we create a grid of points covering Switzerland using the \code{swissRain} object from the \pkg{geostatsp} package. The grid is specified to have 90 cells in the horizontal direction.

\begin{CodeChunk}
\begin{CodeInput}
R> data("swissRain")
R> swissRain = unwrap(swissRain)
R> myRaster = squareRaster(swissRain, 90)
R> dim(myRaster)
\end{CodeInput}
\begin{CodeOutput}
## [1] 62 90  1
\end{CodeOutput} 
\end{CodeChunk} 

In Step 2 we create a matrix with four different parameter sets for the Mat\'ern covariance function. The parameters are denoted as $\kappa$, $\phi$, $\sigma^2$, $\omega$, and $\theta$. 
\begin{CodeChunk}
\begin{CodeInput}
R> params = 
+rbind(c(shape=1.25, range=50*1000, variance = 1.5, 
+        anisoRatio = 1, anisoAngleRadians = 0), 
+        c(2.15, 60*1000, 2,  4, pi/7),
+        c(0.6, 30*1000, 2,  2, pi/5),
+        c(3, 30*1000, 2,   2, pi/7)
+)
\end{CodeInput} 
\end{CodeChunk} 

\begin{CodeChunk}
\begin{CodeInput}
R> params
\end{CodeInput}
\begin{CodeOutput}
##      shape range variance anisoRatio anisoAngleRadians
## [1,]  1.25 50000      1.5          1             0.000
## [2,]  2.15 60000      2.0          4             0.449
## [3,]  0.60 30000      2.0          2             0.628
## [4,]  3.00 30000      2.0          2             0.449
\end{CodeOutput} 
\end{CodeChunk} 

In Step 3, we compute the Mat\'ern covariance matrices using \fct{maternBatch} from \pkg{gpuBatchMatrix}. The returned Mat\'ern covariance matrices are each of size $5580 \times 5580$ and are stacked by row in the output object \code{maternCov}.
\begin{CodeChunk}
\begin{CodeInput}
R> maternCov = gpuBatchMatrix::maternBatch(
+            params, myRaster,          
+            Nglobal=c(256,16), 
+            Nlocal=c(16,4))
R> dim(maternCov)
\end{CodeInput}
\begin{CodeOutput}
## [1] 22320  5580
\end{CodeOutput} 
\end{CodeChunk} 

Step 4 involves performing the Cholesky decomposition on \code{maternCov}. In the \fct{cholBatch} function, the first argument specifies the object to take the Cholesky decomposition. The computed unit lower triangular matrices $L_i$'s are stacked by row in the \code{maternCov} object. The diagonal values of each $D_i$ are returned and stored in each row of the \code{diagMat} object. So if each batch $\Sigma_i$ is of size $n \times n$, then each batch $D_i$ is $1 \times n$.
\begin{CodeChunk}
\begin{CodeInput}
R> diagMat = gpuBatchMatrix::cholBatch(
+          maternCov, 
+          Nglobal= c(512, 16), 
+          Nlocal= c(32, 16), 
+          NlocalCache = 1000)
\end{CodeInput} 
\end{CodeChunk} 

\begin{gather}
\begin{bmatrix} \Sigma_{1} \\ \Sigma_2 \\ \Sigma_3 \\ \vdots
\end{bmatrix}
 \rightarrow
 \begin{bmatrix}
  L_{1} \\ L_{2} \\L_{3} \\ \vdots
  \end{bmatrix} \text{and}
  \begin{bmatrix}
  D_{1} \\ D_{2} \\D_{3} \\ \vdots
  \end{bmatrix}
\end{gather}
In Step 5 we generate two standard Gaussian random vectors, \code{zmatGpu}, which contains $(Z_1, Z_2)$, using the previously created streams with the \fct{clrng::rnormGpu} function. The dimensions of \code{zmatGpu} are specified as \code{c(nrow(maternCov), 2)}.
\begin{CodeChunk}
\begin{CodeInput}
R> zmatGpu = clrng::rnormGpu(c(nrow(maternCov),2), 
+                          streams=streams)
\end{CodeInput} 
\end{CodeChunk} 

Step 6 involves computing $U = L  D^{\frac{1}{2}}  Z$ in batches, utilizing the \fct{multiplyLowerDiagonalBatch} function from \pkg{gpuBatchMatrix}. The matrices \code{maternCov}, \code{diagMat}, and \code{zmatGpu} correspond to the matrices $L$, $D$, and $Z = (Z_1, Z_2)$ respectively in the following illustration:
\begin{CodeChunk}
\begin{CodeInput}
R> simMat = gpuBatchMatrix::multiplyLowerDiagonalBatch(
+                        maternCov, 
+                        diagMat, zmatGpu,
+                        diagIsOne = TRUE,   
+                        transformD = "sqrt", 
+                        Nglobal=c(128, 64, 2), 
+                        Nlocal= c(16, 4, 1), 
+                        NlocalCache=1000)
\end{CodeInput} 
\end{CodeChunk} 

\code{maternCov} (denoted as $L$) is the batched Cholesky decomposition of the Mat\'ern covariance matrices.
\code{diagMat} (denoted as $D$) contains the diagonal values obtained from the Cholesky decomposition.
\code{zmatGpu} (denoted as $Z$) contains the standard Gaussian random vectors $(Z_1, Z_2)$.
\begin{gather}
 \begin{bmatrix}  L_{1} \\ L_{2} \\L_{3} \\ \vdots
 \end{bmatrix} 
 *
  \begin{bmatrix}
   D_{1} \\ D_{2} \\D_{3} \\ \vdots
   \end{bmatrix} 
   *
   \begin{bmatrix}
   Z_{11} & Z_{12}
   \end{bmatrix}
  =
 \begin{bmatrix}
   L_1D_1Z_{11} & L_1D_1Z_{12} \\
   L_2D_2Z_{11} & L_2D_2Z_{12} \\
   L_3D_3Z_{11} & L_3D_3Z_{12} \\
   \vdots  &   \vdots
  \end{bmatrix}
\end{gather}
In Step 7, we convert the computed results $U$ to a spatial \code{raster} object and plot them. 
This step allows us to visualize the simulated Gaussian random fields on a map.
\begin{CodeChunk}
\begin{CodeInput}
R> simRaster = terra::rast(myRaster, nl = ncol(simMat)*nrow(params))
R> values(simRaster) = as.vector(as.matrix(simMat))
\end{CodeInput} 
\end{CodeChunk} 

\begin{figure}[p]
\subfloat[parameter 1, simuation 1\label{fig:maternplot-1}]{\includegraphics[width=0.48\textwidth]{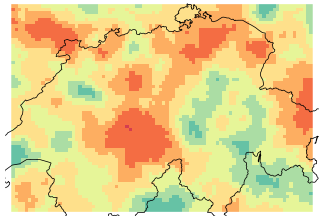} }
\subfloat[parameter 2, simuation 1\label{fig:maternplot-2}]{\includegraphics[width=0.48\textwidth]{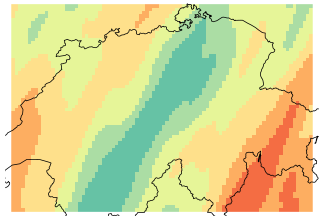} }
\subfloat[parameter 3, simuation 1\label{fig:maternplot-3}]{\includegraphics[width=0.48\textwidth]{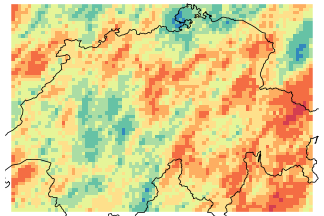} }
\subfloat[parameter 4, simuation 1\label{fig:maternplot-4}]{\includegraphics[width=0.48\textwidth]{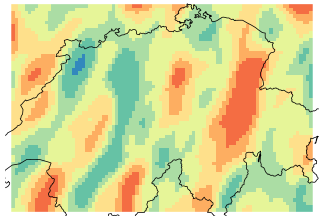} }
\subfloat[parameter 1, simuation 2\label{fig:maternplot-5}]{\includegraphics[width=0.48\textwidth]{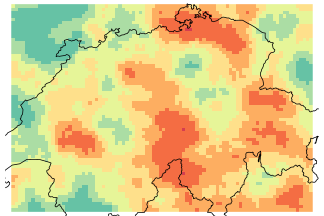} }
\subfloat[parameter 2, simuation 2\label{fig:maternplot-6}]{\includegraphics[width=0.48\textwidth]{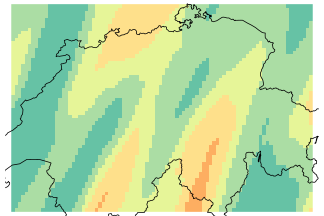} }
\subfloat[parameter 3, simuation 2\label{fig:maternplot-7}]{\includegraphics[width=0.48\textwidth]{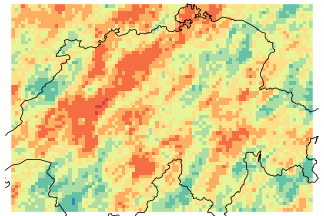} }
\subfloat[parameter 4, simuation 2\label{fig:maternplot-8}]{\includegraphics[width=0.48\textwidth]{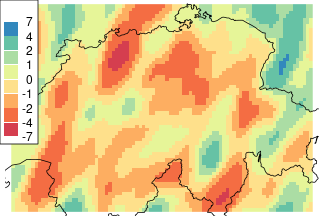} }\caption[Simulated Gaussian random fields]{Simulated Gaussian random fields}\label{fig:maternplot}
\end{figure}

A simple plot can be produced with
\begin{CodeChunk}
\begin{CodeInput}
R> plot(simRaster)
\end{CodeInput} 
\end{CodeChunk} 

and Figure \ref{fig:maternplot} uses some facilities from the \pkg{mapmisc} \citep{mapmiscPackage} package.

\section{Discussion}
The package \pkg{clrng} has been developed to provide \proglang{R} users with access to parallel-generated uniform and some non-uniform random numbers on GPU, facilitating reproducible research in simulations. This includes the ability to set initial seeds in streams on both GPU and CPU, and save and reload the current states of streams. We have applied these GPU-generated random numbers in suitable statistical simulations. For instance, in the Monte Carlo simulation for Fisher’s exact test, for which we are also able to calculate quantiles for the data distribution on the GPU, and in simulating exact Gaussian spatial surfaces. In the latter application, our \pkg{gpuBatchMatrix} package assists by handling the computational work for batches of Mat\'ern covariance matrices, Cholesky decomposition, and matrix multiplication on the GPU. 
Comparisons of performance between using \pkg{clrng} and traditional \proglang{R} on the CPU for some real data examples have shown significant improvements in execution time. These enhancements in computational speed can greatly benefit users in various fields requiring intensive simulations and statistical computations.

By leveraging the \pkg{gpuR} package, \pkg{clrng} provides a user-friendly interface that bridges \proglang{R} and \proglang{OpenCL}. Users can utilize the functionalities in our package without needing to delve into the complexities of \proglang{OpenCL} or even \proglang{\texttt{C++}} code. \pkg{clrng} is portable, as its backend \proglang{OpenCL} supports multiple types of processors. Additionally, it is flexible, as its kernels can be incorporated or adapted in other \proglang{R} packages for diverse applications.


\pkg{clrng} relies on the number and type of \proglang{OpenCL} RNGs used and the capabilities of the \pkg{gpuR} package. As it is closely integrated with the functionalities provided by \pkg{gpuR}, there may be limitations to consider. For example, if developers intend to utilize \pkg{clrng} for tasks not supported within \pkg{gpuR}, such as handling ``sparse'' class objects, they would need to develop custom \proglang{OpenCL} code for these features. 
Additionally, while \pkg{clrng} theoretically operates on any system (Linux, OSX, and Windows) with \proglang{OpenCL} availability, our testing has been conducted only on Ubuntu Linux at the time of writing.

Potential future developments for the \pkg{clrng} package could include: (1) Incorporating additional RNGs, such as the MRG32k3a, LFSR113, and Philox-4×32-10 generators from the \proglang{clRNG} library. Comparative analysis of \proglang{R} performance when using different RNGs on the GPU could also be conducted.
(2) Creating a ``sparse'' matrix class on GPU could enable simulations of Gaussian random fields with sparse correlation structures, such as Gaussian Markov random fields.
(3) Utilizing GPU-generated random numbers in optimized Monte Carlo simulations could significantly enhance performance.

\bibliography{paper1}

\begin{thebibliography}{34}
\newcommand{\enquote}[1]{``#1''}
\providecommand{\natexlab}[1]{#1}
\providecommand{\url}[1]{\texttt{#1}}
\providecommand{\urlprefix}{URL }
\expandafter\ifx\csname urlstyle\endcsname\relax
  \providecommand{\doi}[1]{doi:\discretionary{}{}{}#1}\else
  \providecommand{\doi}{doi:\discretionary{}{}{}\begingroup
  \urlstyle{rm}\Url}\fi
\providecommand{\eprint}[2][]{\url{#2}}

\bibitem[{Allaire \emph{et~al.}(2016)Allaire, Eddelbuettel, Golding, and
  Tang}]{tensorflow1}
Allaire J, Eddelbuettel D, Golding N, Tang Y (2016).
\newblock \emph{tensorflow: R Interface to TensorFlow}.
\newblock \urlprefix\url{https://github.com/rstudio/tensorflow}.

\bibitem[{Bengtsson(2021)}]{future1.19.1}
Bengtsson H (2021).
\newblock \enquote{A Unifying Framework for Parallel and Distributed Processing
  in R using Futures.}
\newblock 10.32614/RJ-2021-048,
  \urlprefix\url{https://journal.r-project.org/archive/2021/RJ-2021-048/index.html}.

\bibitem[{Box(1958)}]{box1958note}
Box GE (1958).
\newblock \enquote{A note on the generation of random normal deviates.}
\newblock \emph{Ann. Math. Statist.}, \textbf{29}, 610--611.

\bibitem[{Brown(2021)}]{mapmiscPackage}
Brown PE (2021).
\newblock \emph{mapmisc: Utilities for Producing Maps}.
\newblock R package version 1.8.0,
  \urlprefix\url{https://CRAN.R-project.org/package=mapmisc}.

\bibitem[{{Determan Jr.}(2017)}]{gpur1}
{Determan Jr} C (2017).
\newblock \emph{gpuR: GPU Functions for R Objects}.
\newblock R package version 2.0.0,
  \urlprefix\url{http://github.com/cdeterman/gpuR}.

\bibitem[{Dyer \emph{et~al.}(1997)Dyer, Kannan, and Mount}]{dyer1997sampling}
Dyer M, Kannan R, Mount J (1997).
\newblock \enquote{Sampling contingency tables.}
\newblock \emph{Random Structures \& Algorithms}, \textbf{10}(4), 487--506.

\bibitem[{Eddelbuettel(2021{\natexlab{a}})}]{cran2021}
Eddelbuettel D (2021{\natexlab{a}}).
\newblock \enquote{CRAN Task View: High-Performance and Parallel Computing with
  R.}
\newblock
  \url{https://cran.r-project.org/web/views/HighPerformanceComputing.html}.
\newblock Version: 2021-11-08.

\bibitem[{Eddelbuettel(2021{\natexlab{b}})}]{eddelbuettel2021parallel}
Eddelbuettel D (2021{\natexlab{b}}).
\newblock \enquote{Parallel computing with R: A brief review.}
\newblock \emph{Wiley Interdisciplinary Reviews: Computational Statistics},
  \textbf{13}(2), e1515.

\bibitem[{Eddelbuettel and Fran\c{c}ois(2011)}]{rcpp1}
Eddelbuettel D, Fran\c{c}ois R (2011).
\newblock \enquote{{Rcpp}: Seamless {R} and {C++} Integration.}
\newblock \emph{Journal of Statistical Software}, \textbf{40}(8), 1--18.
\newblock \doi{10.18637/jss.v040.i08}.
\newblock \urlprefix\url{https://www.jstatsoft.org/v40/i08/}.

\bibitem[{for Health~Statistics(2019)}]{National2018}
for Health~Statistics NC (2019).
\newblock \enquote{Natality 2018. Public use file.}
\newblock Annual internet product. 2019. Available at
  \url{http://www.cdc.gov/nchs/data_access/VitalStatsOnline.htm}.

\bibitem[{Gneiting \emph{et~al.}(2006)Gneiting, {\v{S}}ev{\v{c}}{\'\i}kov{\'a},
  Percival, Schlather, and Jiang}]{gneiting2006fast}
Gneiting T, {\v{S}}ev{\v{c}}{\'\i}kov{\'a} H, Percival DB, Schlather M, Jiang Y
  (2006).
\newblock \enquote{Fast and exact simulation of large Gaussian lattice systems
  in $R^2$: exploring the limits.}
\newblock \emph{Journal of Computational and Graphical Statistics},
  \textbf{15}(3), 483--501.

\bibitem[{Haskard(2007)}]{haskard2007anisotropic}
Haskard KA (2007).
\newblock \emph{An anisotropic Matern spatial covariance model: REML estimation
  and properties.}
\newblock Ph.D. thesis.

\bibitem[{Howes and Thomas(2007)}]{howes2007efficient}
Howes L, Thomas D (2007).
\newblock \enquote{Efficient random number generation and application using
  CUDA.}
\newblock \emph{GPU gems}, \textbf{3}, 805--830.

\bibitem[{Kayibi \emph{et~al.}(2018)Kayibi, Pirzada, and
  Chishti}]{KAYIBI2018298}
Kayibi KK, Pirzada S, Chishti T (2018).
\newblock \enquote{Sampling contingency tables.}
\newblock \emph{AKCE International Journal of Graphs and Combinatorics},
  \textbf{15}(3), 298--306.
\newblock ISSN 0972-8600.
\newblock \doi{https://doi.org/10.1016/j.akcej.2017.10.001}.
\newblock
  \urlprefix\url{https://www.sciencedirect.com/science/article/pii/S0972860017302396}.

\bibitem[{L'Ecuyer(2015)}]{rLEC15a}
L'Ecuyer P (2015).
\newblock \enquote{Random Number Generation with Multiple Streams for
  Sequential and Parallel Computers.}
\newblock In L~Yilmaz, WKV Chan, I~Moon, TMK Roeder, C~Macal, MD~Rossetti
  (eds.), \emph{Proceedings of the 2015 Winter Simulation Conference}, pp.
  31--44. Piscataway, New Jersey: Institute of Electrical and Electronics
  Engineers, Inc.

\bibitem[{L'Ecuyer \emph{et~al.}(2017)L'Ecuyer, Munger, Oreshkin, and
  Simard}]{rLEC17p}
L'Ecuyer P, Munger D, Oreshkin B, Simard R (2017).
\newblock \enquote{Random Numbers for Parallel Computers: Requirements and
  Methods, with Emphasis on {GPUs}.}
\newblock \emph{Mathematics and Computers in Simulation}, \textbf{135}, 3--17.
\newblock Open access at \url{http://dx.doi.org/10.1016/j.matcom.2016.05.005}.

\bibitem[{L'Ecuyer \emph{et~al.}(2021)L'Ecuyer, Nadeau-Chamard, Chen, and
  Lebar}]{rLEC21a}
L'Ecuyer P, Nadeau-Chamard O, Chen YF, Lebar J (2021).
\newblock \enquote{Multiple Streams with Recurrence-Based, Counter-Based, and
  Splittable Random Number Generators.}
\newblock In \emph{Proceedings of the 2021 Winter Simulation Conference}.
\newblock To appear, see
  \url{https://www-labs.iro.umontreal.ca/~lecuyer/myftp/papers/wsc21rng.pdf}.

\bibitem[{L'Ecuyer and Simard(2007)}]{rLEC07b}
L'Ecuyer P, Simard R (2007).
\newblock \enquote{{TestU01}: A {C} Library for Empirical Testing of Random
  Number Generators.}
\newblock \emph{{ACM} Transactions on Mathematical Software}, \textbf{33}(4),
  Article 22.

\bibitem[{L'Ecuyer \emph{et~al.}(2002)L'Ecuyer, Simard, Chen, and
  Kelton}]{l2002object}
L'Ecuyer P, Simard R, Chen EJ, Kelton WD (2002).
\newblock \enquote{An object-oriented random-number package with many long
  streams and substreams.}
\newblock \emph{Operations research}, \textbf{50}(6), 1073--1075.

\bibitem[{L'Ecuyer and Touzin(2000)}]{rLEC00b}
L'Ecuyer P, Touzin R (2000).
\newblock \enquote{Fast Combined Multiple Recursive Generators with Multipliers
  of the Form $a = \pm 2^q \pm 2^r$.}
\newblock In JA~Joines, RR~Barton, K~Kang, PA~Fishwick (eds.),
  \emph{Proceedings of the 2000 Winter Simulation Conference}, pp. 683--689.
  {IEEE} Press.

\bibitem[{Liu \emph{et~al.}(2019)Liu, Li, Sun, and Yu}]{LiuandLi2019}
Liu Y, Li J, Sun S, Yu B (2019).
\newblock \enquote{Advances in Gaussian random field generation: a review.}
\newblock \emph{Computational Geosciences}, \textbf{23}(5), 1011--1047.

\bibitem[{L’Ecuyer(2012)}]{l2012random}
L’Ecuyer P (2012).
\newblock \enquote{Random number generation.}
\newblock In JE~Gentle, WK~Härdle, Y~Mori (eds.), \emph{Handbook of
  computational statistics}, pp. 35--71. Springer, Berlin, Heidelberg.

\bibitem[{L’{E}cuyer \emph{et~al.}(2015)L’{E}cuyer, Munger, and
  Kemerchou}]{l2015clrng}
L’{E}cuyer P, Munger D, Kemerchou N (2015).
\newblock \enquote{{clRNG}: a random number {API} with multiple streams for
  {OpenCL}.}
\newblock Report,
  \urlprefix\url{https://www-labs.iro.umontreal.ca/~lecuyer/myftp/papers/clrng-api.pdf}.

\bibitem[{Mat{\'e}rn(1960)}]{matern1960spatial}
Mat{\'e}rn B (1960).
\newblock \enquote{Spatial variation, Technical Report.}
\newblock \emph{Statens Skogsforsningsinstitut, Stockholm}.

\bibitem[{Mehta and Patel(2011)}]{mehta2011ibm}
Mehta CR, Patel NR (2011).
\newblock \enquote{IBM SPSS exact tests.}
\newblock \emph{Armonk, NY: IBM Corporation}, pp. 23--24.

\bibitem[{Miller and Harrison(2013)}]{10.1214/13-AOS1131}
Miller JW, Harrison MT (2013).
\newblock \enquote{{Exact sampling and counting for fixed-margin matrices}.}
\newblock \emph{The Annals of Statistics}, \textbf{41}(3), 1569 -- 1592.
\newblock \doi{10.1214/13-AOS1131}.
\newblock \urlprefix\url{https://doi.org/10.1214/13-AOS1131}.

\bibitem[{Patefield(1981)}]{patefield1981efficient}
Patefield W (1981).
\newblock \enquote{Algorithm AS 159: An Efficient Method of Generating Random R
  × C Tables with Given Row and Column Totals.}
\newblock \emph{Applied Statistics}, \textbf{30}(1), 91--7.

\bibitem[{{R Core Team}(2021)}]{r2021}
{R Core Team} (2021).
\newblock \emph{R: A Language and Environment for Statistical Computing}.
\newblock R Foundation for Statistical Computing, Vienna, Austria.
\newblock \urlprefix\url{https://www.R-project.org}.

\bibitem[{Ribeiro~Jr. and Diggle(2001)}]{geoR2001}
Ribeiro~Jr P, Diggle P (2001).
\newblock \enquote{{geoR}: a package for geostatistical analysis.}
\newblock \emph{R-NEWS}, \textbf{1}(2), 15--18.
\newblock ISSN 1609-3631.
\newblock \urlprefix\url{http://cran.R-project.org/doc/Rnews}.

\bibitem[{Robert and Casella(2004)}]{robert2004random}
Robert CP, Casella G (2004).
\newblock \enquote{Random variable generation.}
\newblock In \emph{Monte Carlo Statistical Methods}, pp. 35--77. Springer.

\bibitem[{Rupp \emph{et~al.}(2016)Rupp, Tillet, Rudolf, Weinbub, Morhammer,
  Grasser, Jungel, and Selberherr}]{rupp2016viennacl}
Rupp K, Tillet P, Rudolf F, Weinbub J, Morhammer A, Grasser T, Jungel A,
  Selberherr S (2016).
\newblock \enquote{ViennaCL---linear algebra library for multi-and many-core
  architectures.}
\newblock \emph{SIAM Journal on Scientific Computing}, \textbf{38}(5),
  S412--S439.

\bibitem[{Schlather \emph{et~al.}(2015)Schlather, Malinowski, Menck, Oesting,
  and Strokorb}]{RandomFields2015}
Schlather M, Malinowski A, Menck PJ, Oesting M, Strokorb K (2015).
\newblock \enquote{Analysis, Simulation and Prediction of Multivariate Random
  Fields with Package {RandomFields}.}
\newblock \emph{Journal of Statistical Software}, \textbf{63}(8), 1--25.
\newblock \urlprefix\url{http://www.jstatsoft.org/v63/i08/}.

\bibitem[{Sevcikova \emph{et~al.}(2015)Sevcikova, Rossini, and
  L'Ecuyer}]{sevcikova2015package}
Sevcikova H, Rossini T, L'Ecuyer P (2015).
\newblock \enquote{Package ‘rlecuyer’.}
\newblock
  \urlprefix\url{https://cran.r-project.org/web/packages/rlecuyer/index.html}.

\bibitem[{Zhao(2016)}]{zhao_2016}
Zhao P (2016).
\newblock \enquote{R with Parallel Computing from User Perspectives.}
\newblock
  \urlprefix\url{https://parallelr.com/2016/09/10/r-with-parallel-computing/}.

\end{thebibliography}

\newpage

\begin{appendix}

\end{appendix}

\end{document}